\documentclass[12pt]{iopart}
\usepackage{graphicx}
\usepackage{latexsym}
\usepackage{amssymb}
\usepackage{bm}

\newcommand{\f}{\begin{equation}}
\newcommand{\ff}{\end{equation}}

\newcommand{\cg}{\ensuremath{\tilde{\gamma}}}
\newcommand{\ta}{\ensuremath{\tilde{A}}}

\newcommand{\tg}{\ensuremath{\tilde{\Gamma}}}
\newcommand{\tR}{\ensuremath{\tilde{R}}}

\newcommand{\beq}{\begin{equation}}
\newcommand{\eeq}{\end{equation}}
\newcommand{\bea}{\begin{eqnarray}}
\newcommand{\eea}{\end{eqnarray}}

  \def\msol{\ifmmode M_\odot\else$M_\odot$\fi}
  
\begin{document}


\title{Using Gravitational Waves to put limits on Primordial Magnetic Fields}

\author{David Garrison}
\address{Physics Department, University of Houston Clear Lake, Houston, Texas 77058}
\ead{Garrison@UHCL.edu}


\begin{abstract}
We describe a technique for using simulated tensor perturbations in order to place upper limits on the intensity of magnetic fields in the early universe.  As an example, we apply this technique to the beginning of primordial nucleosynthesis.  We determined that any magnetic seed fields that existed before that time were still in the process of being amplified.  In the future, we plan to apply this technique to a wider range of initial magnetic fields and cosmological epochs.
\end{abstract}



\maketitle

\section{Introduction}

Magnetic fields are believed to have played a large part in the dynamics of the evolution of our universe.  However, little is known about the existence of magnetic fields when the universe was very young. There are no direct observations of primordial magnetic fields.  Theories also disagree on the amplitude of primordial magnetic fields. There are currently several dozen theories about the origin of cosmic magnetic fields~\cite{battaner,grasso}.  The main reason that we believe that primordial magnetic fields existed is because they may have been needed to seed the large magnetic fields observed today.   Most theories of cosmic magnetic field generation fall into one of three categories~\cite{battaner,Dolgov:2001,grasso}: 1) magnetic fields generated by phase transitions; 2) electromagnetic perturbations expanded by inflation; and 3) turbulent magnetofluid resulting in charge and current asymmetries.  Once generated, these seed magnetic fields were amplified by a dynamo however, we don't know when or how this dynamo did it's work.

Most models calculate the magnitude of primordial magnetic fields by starting with the observed strength of galactic or intergalactic magnetic fields and calculating how this field should have been amplified or diffused by external effects such as the magnetic dynamo and expansion of the universe~\cite{battaner,grasso}.  A major problem is that there doesn't appear to be a universal agreement of how efficiently a dynamo could have strengthened seed magnetic fields or when the strengthening occurred.  Estimates of the strength of these seed fields can vary by tens of orders of magnitude.  In the absence of  amplification mechanisms, the frozen-in condition of magnetic field lines tells us that \cite{battaner,grasso}.

\f
\vec{B}_0 = \vec{B} a^2.
\ff

Here $\vec{B}_0$ is the present magnetic field where the scale factor is unity and $\vec{B}$  is the magnetic field when the scale factor was a.  Once amplification and diffusion are taken into account, this relationship can be used to calculate the amplitude of magnetic seed fields.  Seed magnetic fields produced during Inflation are predicted to have a current strength somewhere between $10^{-11}$ G and $10^{-9}$ G on a scale of a few Mpc~\cite{battaner,grasso,hoyle}.  Magnetic seed fields generated by phase transitions are believed to be less than $10^{-23}$ G at galactic scales~\cite{battaner,grasso}.  Some turbulence theories imply that magnetic fields were not generated until after the first stars were formed therefore requiring no magnetic seed fields~\cite{battaner}.  

Given how little is understood about primordial magnetic fields and the general lack of agreement among theoretical predictions, it seems clear that the existence of primordial magnetic fields can neither be confirmed or ruled out.  It seems that the best we can do is set an upper limit on the strength of primordial magnetic fields and utilize this limit as a starting point in developing models of cosmic turbulence.  Observations of the CMB limit the intensity of the magnetic seed fields to a current upper limit  of $10^{-9}$ G~\cite{battaner,grasso,hoyle,Kodama}.

It is well known that gravitational waves can interact with a magnetofluid in the presence of a magnetic field.  Work by Duez et al~\cite{duez2} showed how gravitational waves can induce oscillatory modes in a plasma field if magnetic fields are present.  Work by Kahniashvili and others ~\cite{Kahniashvili:2005mp,Kahniashvili:2005qi,Kahniashvili:2006hy,Kahniashvili:2008pf,Kahniashvili:2008pe,Kahniashvili:2009mf} have shown how a turbulent plasma can yield gravitational waves.  The result may be a highly nonlinear interaction as energy is transferred from the fluid to the gravitational waves and back resulting in potentially significant density perturbations.  Magnetic fields are the glue that bind the gravitational waves to the plasma field.  The objective of this work is to utilize the interaction between gravitational waves and the primordial magnetofluid in order to put limits on the strength of magnetic fields that could have existed in the early universe.

\section{Primordial Gravitational Wave Amplitudes}

According to Boyle, Primordial Gravitational Waves develop primarily from tensor perturbations expanded by the inflation event~\cite{Boyle:2005, Boyle:2006}.  The process is similar to that of scalar perturbations and the two are related by a tensor/scalar ratio
\f
r = \Delta^2_t / \Delta^2_s .
\ff
Here the $\Delta$ terms refer to the primordial power spectrums.  As a function of horizon exit time, $\tau_{out}$ and wavenumber, $k$, 
\begin{eqnarray}
\Delta^2_t(k,\tau_{out}) &=& 64 \pi G \frac{k^3}{2 \pi^2} |h_k(\tau_{out})|^2 \approx 8 (\frac{H_{*}}{2 \pi M_{pl}})^2, \\
\Delta^2_s(k,\tau_{out}) &\approx& \frac{1}{2 \epsilon_{*}} (\frac{H_{*}}{2 \pi M_{pl}})^2, \\
r(k) &\equiv & \Delta^2_t(k,\tau_{out}) / \Delta^2_s(k,\tau_{out}) \approx 16 \epsilon_{*}.
\end{eqnarray}
$M_{pl} = (8 \pi G)^{-1/2}$ is the "reduced Planck mass" and $\epsilon$ is the slow roll parameter.  Also the asterisk (*) terms denote the value of the parameters when the tensor perturbation exits the horizon.  The wavenumber is commonly defined as k = $a_{*} H_{*}$ at the horizon exit.  Once the tensor mode enters the horizon, k $\gg$ aH, the strain amplitude of the gravitational waves can be defined as 
\f
h_k = \frac{1}{a \sqrt{2 k}}.
\ff
Unfortunately, because there is not a consistent dimensionless definition of the Hubble Parameter, this method does not allow for an easy way to calculate the amplitude of gravitational waves in the early universe.  We therefore turn to Grishchuk's work \cite{Grishchuk,grishchuk}.  Grishchuk believed that gravitational waves were generated by inflation and amplified by a process called parametric amplification.  Starting with the idea that the gravitational wave power spectrum is deduced by treating contracted tensor perturbations as eigenvalues of a quantum mechanical operator that works on the vacuum state we see that
\f
\left<0\right| \hat{\Omega} \left|0\right> = \left<0\right|h_{ij}(n,\eta)h^{ij}(n,\eta)\left|0\right> =
\frac{\mathcal{C}^{2}}{2\pi^{2}}\int\limits_{0}^{\infty}
~n^{2}\sum_{p=1,2}|{h}_p(\eta)|^{2}\frac{dn}{n}.
\label{meansq}
\ff
Here, n refers to a dimensionless angular wave number, p refers to the left and right handed polarizations of the gravitational waves and $\eta = \int \frac{dt}{a(t)}$ is the conformal time.  The constant $\mathcal{C}$ should be taken as $\mathcal{C}$ = $\sqrt{16 \pi} l_{pl}$.  It can be shown that the mean-square amplitude of the gravitational wave is
\f
h^{2}(n,\eta) = [\frac{4 l_{pl}}{ \sqrt \pi} n ]^{2}\sum_{p=1,2}|{h}_p(\eta)|^{2}.
\ff
The square root of the equation above will provide a root-mean squared (RMS) amplitude of a gravitational wave for a specific wave number.  To complete the power spectrum, we show that the amplitude of gravitational waves can be expressed as
\f
h(n,\eta) = [\frac{4 l_{pl}}{ \sqrt \pi} n ] |{h}(\eta)|.
\ff
Using the relation $n_H = 4 \pi$, which corresponds to the current Hubble radius \cite{Grishchuk,grishchuk}, 
\f
h(n,\eta) = 16 \sqrt {\pi} l_{pl} \frac{n}{n_H} |{h}(\eta)|.
\ff
Grischuk shows that this can be expressed in a convenient form as
\f
h(n) = 16 \sqrt {\pi} l_{pl} \frac{b}{l_0} (\frac{n}{n_H})^{2 + \beta} .
\ff
The variable $\beta$ is the power-law inflation parameter with -2 corresponding to the de Sitter universe and b is a constant defined in terms of $\beta$ as
\f
b = \frac{2^{| 2 + \beta |}} {|1+\beta|^{|1+\beta|}}
\ff
$l_0$ is a constant that denotes an arbitrary Hubble radius during inflation, it is on the order of $10^6~ l_{pl}$ according to Grischuk.  
\f
a(\eta) = l_0 |\eta|^{1+\beta} ~    - \infty \leqslant \eta ~ ;~ -2 \leqslant \beta \leqslant -1
\ff
Since Grishchuk's solution effectively varies by wavenumber, $n_H$, to some power between 0 and -1, we can see that Boyle and Grishchuk's solutions may be equivalent for $\beta = -1.5$.  By setting $n_H$ to reflect a Hubble parameter earlier than the current epoch, we can calculate the spectrum of gravitational waves at any time in the history of the universe post inflation.

\section{Overview of the Software}

As described in the article, Numerical Relativity as a Tool for Studying the Early Universe~\cite{Garrison:2012ex}, the code used here was specifically developed to study relativistic plasma physics in the early universe.  This code is based on the Cactus Framework (www.cactuscode.org).  Cactus was originally developed to perform numerical relativistic simulations of colliding black holes but it's modular design has since allowed it to be used for a variety of Physics, Engineering and Computer Science applications.  It is currently being maintained by the Center for Computation and Technology at Louisiana Sate University.  Cactus codes are composed of a flesh (which provides the framework) and the thorns (which provide the physics).  The code used within this work, SpecCosmo, is a collection of cactus thorns written in a combination of F90, C and C++.

The code uses the relativistic MHD evolution equations proposed by Duez~\cite{duez1}.  It is also designed to utilize a variety of different differencing schemes including 2nd order Finite Differencing, 4th order Finite Differencing and Spectral Methods.  This work uses Fourier Spectral Methods and periodic boundary conditions exclusively.  These involve treating the functions as generic periodic functions and calculating the derivatives using FFTs and inverse FFTs. The code is capable of solving Einstein's Equations directly (through a modified BSSN formulation) as well as the relativistic MHD equations.  The code was thoroughly tested~\cite{Garrison:2012ex} and found to accurately model known GRMHD dynamics.  These tests included MHD waves induced by gravitational waves test, the consistency of cosmological expansion test and shock tests.

The initial data used was derived from work done by several projects involving primordial magnetic fields, phase transitions and early universe cosmology in general~\cite{Durrer:2013pga,islam,Kahniashvili:2008pf,Kahniashvili:2009mf,Kisslinger:2004uc}.  This study models a high energy epoch of the universe after inflation and the Electroweak phase transition when the universe was about 3 minutes old.  The author chose this as the starting point for our study because it was the beginning of the Primordial Nucleosythesis in the early universe.  

\section{Evolution Equations}

The MHD equations used here are based on Duez's evolution equations~\cite{duez1}.

\f
\partial_t \rho_* + \partial_j (\rho_* v^j) = 0 ,
\ff
\f
\partial_t \tilde{\tau} + \partial_i (\alpha^2 \sqrt{\gamma} ~T^{0i} - \rho_* v^i) = s ,
\ff
\f
\partial_t \tilde{S}_i + \partial_j (\alpha \sqrt{\gamma} ~ T^j_i) = \frac{1}{2} \alpha \sqrt{\gamma} ~ T^{ \alpha \beta} g_{\alpha \beta , i} ,
\ff
\f
\partial_t \tilde{B}^i  + \partial_j (v^j \tilde{B}^i - v^i \tilde{B}^j) = 0 .
\ff

Here $\rho_*$ is conserved density, $v^j$ is velocity, $\tilde{\tau}$ is the energy variable, $\tilde{S}_i$ is the momentum variable, s is the source term, $\alpha$ is the lapse term, $\gamma$ is the determinate of the three metric and $T^{ij}$ is the stress-energy tensor.  The tilde denotes that the term was calculated with respect to the conformal metric.  The first equation comes from conservation of baryon number, the second derives from conservation of energy, the third is conservation of momentum and the fourth is the magnetic induction equation.  For this simulation we use Geodesic Slicing, $\alpha$ = 1.0, $\beta_i$ = 0.0.  

The code utilizes a first order version of the BSSN equations to simulate the background space-time.  For fixed gauge conditions, the modified BSSN equations as defined by Brown~\cite{Brown:2012me} are:

\begin{eqnarray}
\overline\partial_0 K &=&
 \alpha\left( \ta^{ij}\ta_{ij} + \frac{1}{3}\, K^2 \right) + 4 \pi \alpha ( \rho + S) \ .\label{eq:K-evol} \\
\overline\partial_0 \phi &=& -\frac{\alpha}{6}\, K  \ ,\label{eq:phi-evol}
\\
\overline\partial_0 \phi_i &=& -\frac{1}{6}\alpha \overline D_i K - \kappa^\phi {\cal C}_i \ ,
\\
\overline\partial_0 \cg_{ij} &=& -2\alpha\ta_{ij} \ ,\label{eq:gammatilde-evol}
\\
\overline\partial_0 \ta_{ij} &=& e^{-4\phi}\left[
 \alpha(\tR_{ij} - 8 \pi S_{ij})- 2\alpha \overline D_{(i}\phi_{j)} + 4\alpha \phi_i\phi_j 
+ \Delta\tilde\Gamma^k{}_{ij}(2\alpha\phi_k) \right]^{TF} 
\nonumber\\ 
 & & + \alpha K\ta_{ij} - 2\alpha\ta_{ik}\ta^k_{\; j} \ ,\label{eq:Atilde-evol}
\\
\overline\partial_0 \tilde\gamma_{kij} &=& -2\alpha\overline D_k\ta_{ij}
 - \kappa^\gamma  {\cal D}_{kij} \ ,\label{eq:dkijtilde-evol}
\\
\overline\partial_0 \tilde\Lambda^i &=& - \frac{4}{3}\alpha \tilde D^i K
 + 2\alpha\left( \Delta\tg^i{}_{k\ell}\ta^{k\ell} + 6\ta^{ij} \phi_j - 8 \pi  \cg^{ij} S_j \right) \ .
\label{eq:gamma-evol}
\end{eqnarray}

The bar denotes a derivative taken with respect to the fiducial metric and the tilde again denotes a derivative taken with respect to the conformal metric.  Also, ${\cal C}_i$ and ${\cal D}_{kij}$ are constraint equations and $\kappa^\phi$ and $\kappa^\gamma$ are proportionality constants.  $\rho$, $S$, $S_j$ and $S_{ij}$ are source terms as found in the standard version of the BSSN equations.  Brown et al also defined:

\begin{eqnarray*}
{\cal C}_i &=& \phi_i - \overline D_{i}\phi = 0 , \\
{\cal D}_{kij} &=&  \tilde{\gamma}_{k i j} - \overline{D}_k \tilde{\gamma}_{ij} = 0 , \\
\Delta\tilde{\Gamma}^i{}_{k\ell} &=& \frac{1}{2}\tilde{\gamma}^{ij}\left(
 \tilde{\gamma}_{k\ell j} + \tilde{\gamma}_{\ell kj} - \tilde{\gamma}_{jk\ell} \right) \ , \nonumber\\
    \tR_{ij}  & = & -\frac{1}{2} \tilde\gamma^{k\ell} \overline D_k \tilde\gamma_{\ell ij} 
      + \tilde\gamma_{k(i} \overline D_{j)} \tilde\Lambda^k 
		+ \tilde\gamma^{\ell m}\Delta\tilde\Gamma^k_{\ell m} \Delta\tilde\Gamma_{(ij)k} \\
		& & + \tilde\gamma^{k\ell} [ 2\Delta\tilde\Gamma^m{}_{k(i} \Delta\tilde\Gamma_{j)m\ell} 
			+ \Delta\tilde\Gamma^m{}_{ik} \Delta\tilde\Gamma_{mj\ell} ] 
		  \ ,
\nonumber\\
\end{eqnarray*}

\section{Experimental Set-up and Assumptions}

In order to determine the upper limit of primordial magnetic fields that existed at a particular stage in the evolution of our universe, we inject a broad spectrum of gravitational waves into a homogenous relativistic plasma field with a constant magnetic field and study the results.  This is similar to what Duez did in the second of two papers~\cite{duez2}.  The basic idea of the Duez paper was to calculate the effect that standing gravitational plane waves would have on a homogenous plasma field with a constant magnetic field.  The result was to excite magnetosonic and Alfen waves in the plasma based on the polarization of the gravitational waves and other parameters such as the density, temperature and magnetic field of the plasma.  This was done as a test of their GRMHD code but we use it here to probe what magnetic fields may have been physically allowable in the early universe.  We choose to perform this study 180s after the big bang although such a study could have been performed anytime after electro-weak symmetry breaking.  For the results to be relevant, we must assume that magnetogenesis and any dynamo effects had already created and strengthened a primordial magnetic fields that would gradually be weakened by the expansion of the universe.  The temperature, density, Hubble parameter, scale factor and mass contribution of the universe at this stage are all well known~\cite{islam}.  We utilized an initial temperature of $1.0 \times 10^{9}$ K.  The scale factor and Hubble Parameter are $a = 2.81 \times 10^{-9}$ and $H = 2.46 \times 10^{-3} s^{-1}$ respectively.  The mass/energy density at the time was $1.08 \times 10^{4} \frac{kg}{m^3}$. Our study assumes that 80\% of the mass density of the universe was composed of "dark matter".  This was chosen to be consistent with our current dark matter to baryonic matter ratio.  This "dark matter" was simulated using a pressureless, non-magnetic fluid with no internal energy, in addition to the magnetofluid used to simulation regular matter.  This was done to keep us from over estimating the effects of magnetic fields on the matter field.  The amplitude of the gravitational waves at this epoch was determined using Grishchuk's solution described in a previous section.  

\begin{figure}  
\begin{center}
\includegraphics[height=3.5in,width=4.50in,angle=0]{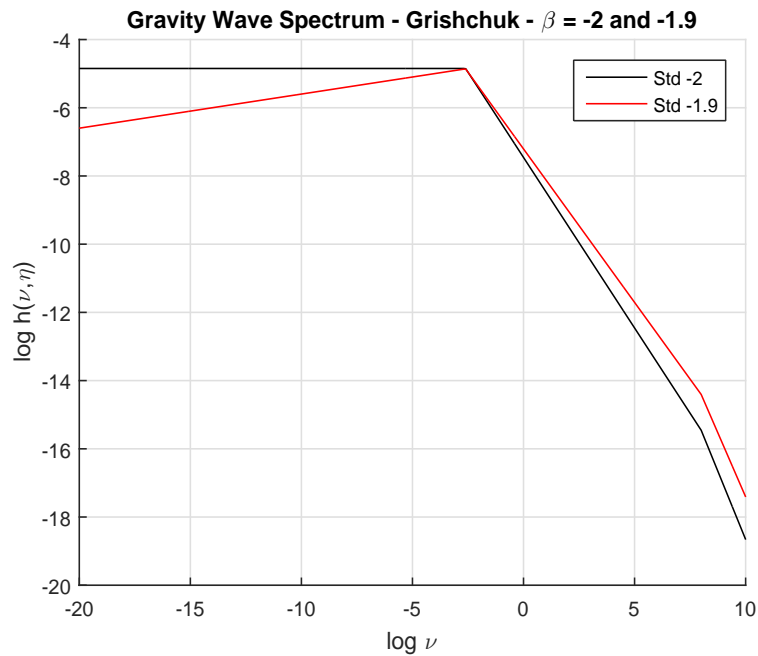}
\caption{\small \sl Primordial Gravitational Wave spectrum as calculated by Grishchuk's method for t = 180 s.}  
\end{center}
\end{figure} 

We ran 6 simulations with different values of a fixed magnetic fields along the z-axis, $0~G$, $10^{2}~G$, $10^{4}~G$, $10^{5}~G$, $10^{6}~G$ and $10^{8}~G$.  Each run used random tensor perturbations with amplitudes up to $10^{-19}$.  We utilized a three dimensional computational grid with $64^3$ internal grid points corresponding to $4^3$ meters with a courant factor of 0.1.  The domain size of $4^3$ meters was chosen to allow for multiple light crossing times during the course of the simulation.  Geodesic slicing conditions and periodic boundary conditions were used for all simulation runs.  We also used a 3rd order Iterative Crank Nicolson time scheme for time integration.  A spectral differencing method was used and the simulations ran for over 1,000 iterations.  There were no shocks or discontinuities in the system so we did not utilize our HRSC routines. 

\section{Results}

As one can see from Figure 2, the density perturbations appear larger as the intensity of the initial magnetic field increases.  There appears to be no difference between the 0 G magnetic field and the $10^4$ G magnetic field.  However, the $10^{5}$ G magnetic field seems to have a much larger effect on the plasma field with density perturbations on the order of a part in $10^{12}$ result.  When the magnetic fields are near or above $10^{5}$ G the perturbations continue to grow until the system becomes unstable.  This is clearly an unphysical result.  It should be noted that a primordial magnetic field of $10^{8}$ would correspond to a current cosmological magnetic field of $10^{-9}$ G which is the established upper limit.

\begin{figure}  
\begin{center}
\includegraphics[height=3.5in,width=4.50in,angle=0]{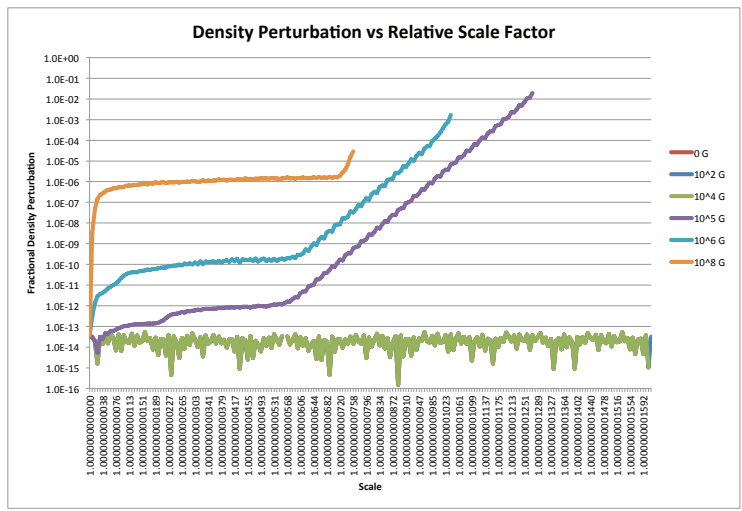}
\caption{\small \sl Density perturbations as the result of different initial magnetic fields.}  
\end{center}
\end{figure}

\section{Discussion}

The goal of this project was to develop a technique for testing the upper limit of cosmological magnetic fields throughout different epochs of universal evolution.  We did this using the beginning of Primordial Nucleosythesis as an example.  We observed that the the relative amplitude of density perturbations varied according to the strength of the initial magnetic fields.  Our observed instabilities for magnetic fields greater than $10^4$ G imply that such strong magnetic fields should not have been physically possible during the Primordial Nucleosynthesis epoch.  We saw that the maximum possible magnetic field as determined by observation, is not physically viable.  From this we conclude that the amplification of the seed magnetic fields either did not finish until much later or current cosmological magnetic fields should have amplitudes below $10^{-13}$ G.  Future work will involve applying this technique to later epochs over a wider range of initial magnetic fields in order to more accurately determine upper limits for magnetic field intensities. 

\section{Conflicts of Interests}

The author declares that there are no conflict of interests regarding the publication of this article.

\section{Acknowledgement}

The author would like to acknowledge the support of the University of Houston Center for Advanced Computing and Data Systems for access to the high performance computing resources used for the completion of this project. 


\end{document}